\documentclass[twocolumn,showpacs,amsmath,amssymb, nobibnotes, aps, prl,showkeys]{revtex4}
\usepackage{graphicx}
\usepackage{dcolumn}
\usepackage{bm}
\usepackage{docs}
\usepackage{boldline,multirow}
\usepackage[dvipsnames]{xcolor}
\expandafter\ifx\csname package@font\endcsname\relax\else
 \expandafter\expandafter
 \expandafter\usepackage
 \expandafter\expandafter
 \expandafter{\csname package@font\endcsname}%
\fi
\newcommand{\be}{\begin{equation}}
\newcommand{\ee}{\end{equation}}
\newcommand{\bea}{\begin{eqnarray}}
\newcommand{\eea}{\end{eqnarray}}

\begin{document}
\title{Implications of a proton blazar inspired model on correlated observations of neutrinos with gamma-ray flaring blazars}

\author{Prabir Banik$^{1}$\thanks{Email address: pbanik74@yahoo.com}, Arunava Bhadra$^{2}$\thanks{Email address: aru\_bhadra@yahoo.com}, Madhurima Pandey$^3$\thanks{Email address: madhurima.pandey@saha.ac.in} and Debasish Majumdar$^3$\thanks{Email address: debasish.majumdar@saha.ac.in}}
\affiliation{ $^1$Surendra Institute of Engineering $\&$ Management, Dhukuria, Siliguri, West Bengal, India 734009}
\affiliation{ $^{2}$High Energy $\&$ Cosmic Ray Research Centre, University of North Bengal, Siliguri, West Bengal, India 734013}
\affiliation{ $3$ Astroparticle Physics and Cosmology Division, Saha Institute of Nuclear Physics, HBNI 1/AF Bidhannagar, Kolkata, West Bengal, India 700064}
\begin{abstract}

The recent detection of the neutrino events IceCube-170922A, 13 muon-neutrino events observed in 2014-2015 and IceCube-141209A by IceCube observatory from the Blazars, namely TXS 0506+056, PKS 0502+049/TXS 0506+056 and GB6 J1040+0617 respectively in the state of enhanced gamma-ray emission, indicates the acceleration of cosmic rays in the blazar jets. The photo-meson ($p\gamma$) interaction cannot explain the IceCube observations of 13 neutrino events. The non-detection of broadline emission in the optical spectra of the IceCube blazars, however, question the hadronuclear (pp) interaction interpretation through relativistic jet meets with a high density cloud. In this work, we investigate the proton blazar model in which the non-relativistic protons that come into existence under the charge neutrality condition of the blazar jet can offer sufficient target matter for $pp$ interaction with shock-accelerated protons, to describe the observed high-energy gamma-rays and neutrino signal from the said blazars. Our findings suggest that the model can explain consistently the observed electromagnetic spectrum in combination with the appropriate number of neutrino events from the corresponding blazars.

\end{abstract}

\maketitle

\section{Introduction}
Recently the IceCube Neutrino Observatory at the South Pole reported the detection of a few reconstructed high energy (TeV energy and above) neutrino events in spatial coincidence with a couple of known gamma-ray blazars which provide the first direct identification of sources of high energy cosmic rays. Gamma-ray blazars, a class of active galactic nuclei (AGN) with powerful relativistic jets oriented close to the line of sight of the observer, are considered as one of the promising contributors to the diffuse flux of high-energy neutrinos detected by IceCube \cite{Biermann87}. Blazars are usually sub-classified into BL Lac objects and flat spectrum radio quasars (FSRQs), depending on the emission line properties \cite{Urry95}. A common feature of the nonthermal electromagnetic (EM) spectral energy distribution (SED) of blazars is the double-hump structure $-$ one at the IR/optical/UV or X-ray and the other at high energy gamma-ray bands. The lower-energy bump is usually believed to produce from synchrotron radiation of primary electrons while the second one can be explained by several different mechanisms. The most popular explanation of higher-energy hump lies in inverse Compton (IC) scattering of synchrotron or external photons \cite{Ghisellini85,Ghisellini09,Sikora94}. The leptonic scenarios, however, cannot explain some observed characteristics such as very fast variability almost in all observed bands \cite{Albert07,Aharonian07}. 

The first detection of cosmic ray sources occurs on 22 September, 2017 when the IceCube collaboration observed a high-energy muon-neutrino event IceCube-170922A of energy $\sim 290$ TeV  \cite{IceCube18a, IceCube18b} coming from the direction of the sky location of the known blazar TXS 0506+056, a BL Lac object \cite{IceCube18a,Ansoldi18}. A follow-up observation by the Fermi Large Area Telescope (LAT) Collaboration \cite{Tanaka17} revealed that the gamma-ray source TXS 0506+056 blazar was in a state of enhanced emission in GeV energies with day-scale variability \cite{Keivani18} on September 28, 2017. A significant very-high-energy $\gamma$-ray signal has been observed by the Major Atmospheric Gamma Imaging Cherenkov (MAGIC) Telescopes \cite{Mirzoyan17} with energies up to about 400 GeV on 28 September 2017. As high energy neutrinos are believed to be produced only in hadronic processes, the observed association of the neutrino event with the gamma-ray flaring blazar TXS 0506+056 has opened a new window to study the origin of cosmic rays in blazars using multi-messenger astronomy. 

Triggered by the discovery of the 2017 flare from TXS 0506+056 blazar, IceCube Collaboration re-investigated the archival 9.5 yr of IceCube data at the position of TXS 0506+056 and reported significant evidence for a flare of 13 muon-neutrino events during September 2014 and March 2015 \cite{IceCube18a}. Assuming a power law distribution of the signal between 32 TeV and 3.6 PeV energy range, a statistical significance $3.5\sigma$ excess over the atmospheric neutrino background was found during a 158-day box-shaped time window from MJD 56937.81 to MJD 57096.21 \cite{IceCube18a}. Surprisingly, at the arrival time window of such a neutrino flare, the blazar TXS 0506+056 was found to be in the quiescent state of both the radio and GeV emission \cite{Padovani18}. A nearby FSRQ blazar, PKS 0502+049 which is only $\sim 1.2^0$ far from TXS 0506+056, was in a state of enhanced gamma-ray emission state just before and after the period of the neutrino excess in 2014$-$2015 \cite{Padovani18,He18}. Thus in principle, the production of neutrinos in the jet of PKS 0502+049 blazar could make some contribution to such neutrino flare in $2014-2015$ as the position of PKS 0502+049 is found to be spatially consistent the directional reconstruction uncertainties of such observed muon neutrinos.

Very recently, the IceCube Observatory reported the detection of a reconstructed high energy neutrino event, designated as IceCube-141209A \cite{Garrappa19} in spatial coincidence with another known gamma-ray blazar, GB6 J1040+0617 with the coincidence detection probability by chance is just 30\%. The observed association of the neutrino suggests that the blazar GB6 J1040+0617 be another plausible neutrino source candidate.

All those correlated observations of the high-energy neutrinos with blazars during a gamma-ray flaring stage revealed that blazars may indeed be one of the most possible extragalactic sources of very-high-energy cosmic rays. A relevant question is the production scenario of detected neutrinos and gamma rays from the blazars. The high energy neutrinos can be produced either in lepto-hadronic ($p\gamma$) or in pure hadronic (pp) interactions. The high energy neutrinos and TeV gamma rays (totally or partially) are produced in the former scenario through the interaction of blazar accelerated cosmic rays with surrounding EM radiation whereas in the latter scenario they are produced in the interaction of the blazar accelerator cosmic rays with the ambient matter. Another issue is the maximum energy of the accelerated particles in the detected sources that led to the creation of such high energy neutrinos together with observed EM radiation from the sources. 

In previous work, it was demonstrated that a proton Blazar model can describe consistently the observed high energy gamma rays and neutrino signal from the blazar TXS 0506+056 \cite{Banik19}. In the present work, we would like to demonstrate that the proton blazar model can consistently explain the spectral behavior of observed higher-energy bump of the EM SED along with the observed association of neutrinos from all the three IceCube blazars assuming that the association of the observed neutrino events with the corresponding blazars at the flaring stage are genuine. Such a scenario appears to be more realistic than the scenario of the cloud-in-jet model as we discuss later. 

The plan of the paper is the following: In the next section, we briefly review the models proposed in the literature to explain the observed high energy gamma rays and neutrino signal from the Icecube blazars. The proton-Blazar model will be described in section III. In the same section, we shall describe the methodology for evaluating the gamma-ray and neutrino fluxes produced in the interaction of cosmic rays with the ambient matter in the AGN jet under the framework of proton blazar model.  The numerical estimated fluxes of hadronically produced gamma-rays and neutrinos from the Icecube blazars over the GeV to TeV energy range will be shown in Sec. IV along with the observations. The findings of the present work will be discussed in Sec. V and we shall conclude finally in Sec. VI.

\section{The models of gamma rays and neutrinos production in Icecube blazars}
Several efforts have been made so far to interpret the production of the detected neutrino events together with the EM observations from TXS 0506+056. A common feature of all the proposed models is that protons are accelerated along with electrons to relativistic energies in the acceleration sites. The observations of the HE component have been interpreted mainly with the interaction of the protons either with low energy photons of blazar environment (lepto-hadronic ($p\gamma$) interaction)\cite{Ansoldi18,Keivani18,Gao19,Cerruti19} and/or with ambient matter (hadronic ($pp$)) \cite{Liu19,Sahakyan18}. 

Ansoldi et al. (2018) \cite{Ansoldi18} described the detected neutrino event along with the EM observations from the said blazar by assuming a dense field of external low-energy photons originating in a possible structured-layer surrounding the jet as targets for photohadronic interactions. Keivani et al. (2018) \cite{Keivani18} assumed a hybrid leptonic scenario of the blazar TXS 0506+056 where the production of high-energy gamma-rays was described by external inverse-Compton processes and high-energy neutrinos are accounted via a radiatively sub-dominant hadronic component. The said observation was interpreted recently in Gao et al. (2019) \cite{Gao19} by involving a compact radiation core for high photohadronic interaction rates. 

In the hadronic ($pp$) interaction scenario, the high thermal plasma density is required for efficient high energy $\gamma-$ray production in AGN jet. Recently, Liu et al. (2019) \cite{Liu19} described the observed EM and neutrino fluxes from the blazar TXS 0506+056 by assuming the presence of clouds in the vicinity of the super-massive black hole (SMBH) that provides targets for inelastic $pp$ collisions once they enter the jet. However, the non-detection of the BLR emission from TXS 0506+056 and other BL Lac objects \cite{Keivani18} create doubt on the presence of broadline region (BLR) clouds in the vicinity of the SMBH of TXS 0506+056 \cite{Keivani18}. Murase et al \cite{mur18} considered the CR-induced neutral beam model in which beamed neutrons, which are produced via the photodisintegration of nuclei in the blazar zone, interact with an external radiation field/cloud after escaping from the blazar zone and thereby produce neutrinos. Their model can naively explain both of the 2017 and 2014-2015 neutrino
flares of TXS 0506+056 when effective optical depth to the photodisintegration process is taken $0.1$ and $\ge 1$ respectively. 

The main reason behind the difficulties to understand the interaction mechanism for gamma-ray and neutrino production is the composition of the bulk of the jet medium which is not clearly known. For high luminous blazars, the proton component of plasma is necessary as suggested by some author e.g, Celotti \& Fabian (1993) \cite{Celotti93}, Ghisellini et al. (2010) \cite{Ghisellini10}, in order to maintain the radiated power which would not exceed that carried by jet. Under such a scenario, recently two of us (Banik and Bhadra) have demonstrated that the detected neutrino event together with the EM observations from TXS 0506+056 can be consistently described by assuming a proton blazar model where non-relativistic protons that come into existence under the charge neutrality condition of the blazar jet can offer sufficient target matter for $pp$ interaction with shock-accelerated protons \cite{Banik19}. 

So far no proper explanation of the Icecube observed flare of 13 muon-neutrino events is available in the literature. Rodrigues et al. (2019)\cite{Rodrigues19} concluded from their analysis that the high event numbers of neutrinos quoted by IceCube can not be explained by any other model of the source. Considering a EM spectral hardening of the source TXS 0506+056 above 2 GeV during the neutrino flare, as predicted by Padovani et al. (2018) \cite{Padovani18} based on Fermi data, Rodrigues et al. (2019) \cite{Rodrigues19} recently demonstrated that hardly two to five neutrino events during the flare can be described with different leptohadronic models: a one-zone model, a compact-core model, and an external radiation field model. Garrappa et al. (2019) \cite{Garrappa19} pointed out that the feature of spectral hardening in SED of the source may in fact not be significant. On the other hand, Liang et al. (2018) \cite{Liang18} and He et al. (2018) \cite{He18} have interpreted the 2014 neutrino flare and the gamma-ray flare with a jet-cloud interaction model assuming that the 2014 detection was actually from the nearby source PKS 0502+049.

No detail production model for neutrinos from GB6 J1040+0617 is available in the literature yet. 

\section{The proton-blazar model and computation technique for flux estimation}
The overall composition of the AGN jet is an unknown issue so far. It is generally assumed that the relativistic jet material is composed of relativistic protons (p) and electrons ($e^{-}$) in almost all hadronic models of AGN jet. In principle, cold (non-relativistic) protons that arose from charge neutrality condition, also exist as described in the adopted proton blazar inspired model \cite{Banik19} which is developed from the proton-blazar model \cite{Mucke01}. The existence of cold protons is supported by the fact that only a small fraction of protons of the system (roughly $4 \%$) are accelerated to non-thermal energies at diffusive shocks as demonstrated from hybrid simulation study \cite{cap15}.  The adjustable parameters of the model are the ratio of the number of relativistic protons to electrons, the maximum energies attained by protons/electrons in acceleration process and slope of their energy spectrum, luminosities of electrons and protons.  We consider that the region of a blazar jet which is responsible for the non-thermal emission, is a spherical blob of size $R_b'$ (primed variables for jet frame) and it contains a tangled magnetic field of strength $B'$. The magnetic field energy density can be written as $u_B' = B'^2/8\pi = 3 p_B'$ where $p_B'$ is the corresponding pressure. If $\theta$ be the angle between the line of sight and the jet axis then the Doppler factor of the moving blob can be written as $\delta = \Gamma_j^{-1}(1-\beta_j\cos\theta)^{-1}$ where $\Gamma_j = 1/\sqrt{1-\beta_j^2}$ is the bulk Lorentz factor \cite{Petropoulou15}. 

In the proton blazar framework, we have assumed a broken power law energy distribution of accelerated relativistic electrons in the blazar jet to explain the low-energy bump of the SED by synchrotron radiation and can be written as \cite{Katarzynski01,Banik19}
\begin{eqnarray}
N_e'(\gamma_e') = K_e \gamma_e'^{-\alpha_1} \hspace{1.5cm} \mbox{if}\hspace{0.6cm} \gamma_{e,min}' \le \gamma_e' \le \gamma_b' \nonumber \\
         = K_e \gamma_b'^{\alpha_2-\alpha_1} \gamma_e'^{-\alpha_2} \hspace{0.35cm} \mbox{if}\hspace{0.46cm} \gamma_b' <\gamma_e' \le \gamma_{e,max}'\;
\label{Eq:1}
\end{eqnarray}
where $\alpha_1$ and $\alpha_2$ are the spectral indices before and after the spectral break at Lorentz factor $\gamma_b'$ and $\gamma_e' = E_e'/m_e c^2$ is the Lorentz factor of electrons of energy $E_e'$. The normalization constant $K_e$ can be found using the relation \cite{Bottcher13,Banik19}
\begin{equation}
L_e' = \pi R_b'^2 \beta_j c \int_{\gamma_{e,min}'}^{\gamma'_{e,max}}m_e c^2\gamma_e' N_e'(\gamma_e') d\gamma_e'
\label{Eq:2}
\end{equation}
where $L_e'$ represents the kinetic power of accelerated electrons in the co-moving blazar jet frame. The energy density and number density  of relativistic (`hot') electrons are $u_e'= \int m_e c^2\gamma_e' N_e'(\gamma_e') d\gamma_e'$ and $n_{e,h}' = \int N_e'(\gamma_e') d\gamma_e'$ respectively. 

It is often considered that all the electrons in a concerned system undergo Fermi acceleration. However, as pointed out by Eichler and Waxman \cite{Eichler05} in the context of gamma ray bursts that the exact fraction of electrons ($\chi_e$) participated in diffusive shock (Fermi) acceleration cannot be evaluated by current observations; the observationally admissible range is $m_e/m_p \le \chi_e \le 1$. When thermal ions/electrons encounter at any shock barrier, only about $ 25 \;\%$ of them are reflected. When thermal ions/electrons impinge a shock barrier too weak to reflect them, they cross toward downstream and thus do not participate in the acceleration. A part of the impinged ions/electrons are reflected by shocks and  energized up to a certain level via diffusive shock acceleration (DSA) also finally avected downstream.  Only a small fraction of injected  ions/electrons achieve sufficient energy via DSA and escape toward upstream \cite{cap15}. In the present work we have taken $\chi_e \approx 10^{-3}$, which is within the allowed range and consistent with the hybrid simulation results of DSA by parallel collisionless  shock \cite{Gia92, Bykov96}. Such a low value is also supported by the fact that electrons share nearly two order less total energy compare to total energy of protons \cite{Vazza15, Bykov96} and as mentioned already that only $0.04$ fraction of protons undergo Fermi acceleration. Note that it is not mandatory to strictly consider such low $\chi_e \sim 10^{-3}$, we may also opt up to $\chi_e \sim 10^{-2}$. In the latter case we have to increase the cosmic ray flux appropriately i.e. we need higher jet power. The total number electrons including `hot' and non-relativistic (`cold') electrons is $n_{e}' = n_{e,h}'/\chi_e$.  

The detailed computation technique of electromagnetic and neutrino spectra at the Earth from a blazar in the framework of proton-blazar inspired model is given in our earlier work \cite{Banik19}. Basically the parameters are chosen such a way that synchrotron emission of the relativistic electrons gives the low energy component of the EM SED of the blazar which is computed here following the formulation given in \cite{Bottcher13}.


The inverse Compton (IC) scattering of primary accelerated electrons with the seed photons co-moving with the AGN jet is employed to describe the lower part of the high energy component of EM SED of the blazar. The  emissivity $Q_{c}(\epsilon_{c}')$ of produced gamma-ray photons of energy $E_{c}'$ ($= m_e c^2 \epsilon_{c}'$ due to IC scattering of primary accelerated electrons with the seed photons is given in \cite{Blumenthal70,Inoue96}. The seed photon density and spectra are estimated directly from the observed photon flux from the blazar \cite{Banik19, Dermer02}. 

In the proton blazar framework, we have assumed a power law behavior of the cosmic ray protons which are supposed to be accelerated to very high energies $E_p' = m_p c^2 \gamma_p'$ in the blob of a blazar jet \cite{Malkov01, Cerruti15}

\begin{equation}
 N_p'(\gamma'_p) =  K_p {\gamma'_p}^{-\alpha_p} .
\end{equation}
where $\gamma_p'$ is the Lorentz factor of accelerated protons, $\alpha_p$ represents the spectral index, $K_p$ indicates the proportionality constant which can be obtained from Eq. (2) (as the expression also holds for proton) using the corresponding jet power $L_p'$ for relativistic protons. The energy density of relativistic protons is $u_p' = \int m_p c^2\gamma_p' N_p'(\gamma_p') d\gamma_p'$ and $n_p' = \int N_p'(\gamma_p') d\gamma_p' $ represents the corresponding number density of relativistic protons.

Secondary particles (mainly pions) are produced when the shock accelerated cosmic rays interact with the cold matter (protons) of density $n_{H} = (n_{e}'-n_p')$ in the blob of AGN jet. The emissivity of secondary particles has been calculated in this work following \cite{Liu19,Anchordoqui07,Banik17a,Kelner06}.

The $\pi^{0}$ mesons subsequently decays to gamma-ray . The produced TeV$-$PeV gamma-rays is likely to be absorbed due to internal photon-photon ($\gamma\gamma$) interactions \cite{Aharonian08} while propagating through an isotropic source of low-frequency radiation which is generally assumed to be the observed synchrotron radiation photons produced by the relativistic electron population in the co-moving jet.
The gamma-ray emissivity as a function of gamma-ray energy $E_{\gamma}'( = m_e c^2 \epsilon_{\gamma}')$ has been computed following \cite{Banik17b} and the emissivity of escaped gamma-rays after internal $\gamma\gamma$-absorption within the source region is estimated following \cite{Bottcher13}. 

\begin{eqnarray}
Q_{\gamma,esc}'(\epsilon_{\gamma}') = Q_{\gamma}'(\epsilon_{\gamma}') .\left( \frac{1-e^{-\tau_{\gamma \gamma}}}{\tau_{\gamma \gamma}} \right).
\end{eqnarray}
where $\tau_{\gamma \gamma}(\epsilon_{\gamma}')$ is the optical depth for the interaction and can be obtained as given by \cite{Aharonian08, Banik19}. 

The total number of high-energy injected electrons/positrons ($Q_{e}'$) in the emission region of AGN jet are the sum of those created in $\gamma\gamma$ pair production and those produced directly due to the decay of $\pi^{\pm}$ mesons created in $pp$ interaction. These injected electrons/positrons will initiate EM cascades in the AGN blob via the synchrotron radiation and the IC scattering. The secondary pair cascading processes were executed following the self-consistent formalism of B$\ddot{o}$ttcher et al. (2013) \cite{Bottcher13} after inclusion of IC mechanism.

The observable differential flux of gamma-rays reaching the earth from a blazar can be written as

\begin{eqnarray}
E_{\gamma}^2\frac{d\Phi_{\gamma}}{dE_{\gamma}} = \frac{V'\delta^2\Gamma_j^2}{4\pi d_L^{2}}\frac{E_{\gamma}'^2}{m_e c^2} Q_{\gamma,esc}'(\epsilon_{\gamma}') . e^{-\tau_{\gamma\gamma}^{EBL}}
\label{eqgamflux}
\end{eqnarray}
where $Q_{\gamma,esc}'(\epsilon_{\gamma}')$ is the total gamma-ray emissivity from the blob of AGN jet with photon energies $E_{\gamma}' = m_e c^2 \epsilon_{\gamma}'$ in co-moving jet frame including all processes stated above i.e, the synchrotron and the IC radiation of accelerated electrons, the gamma-rays produced in $pp$ interaction and also the synchrotron photons of EM cascade electrons, $V' = \frac{4}{3}\pi R_b'^3$ is the volume of the emission region, $E_{\gamma} = \delta E_{\gamma}'/(1+z) $ \cite{Atoyan03} relates photon energies in the observer, $d_L$ is the luminosity distance between the AGN and the Earth and co-moving jet frame of red shift parameter $z$ respectively. Here we introduced the effect of the absorption by the extragalactic background (EBL) light on gamma-ray photons and $\tau_{\gamma\gamma}^{EBL}(\epsilon_{\gamma},z)$ is the corresponding optical depth which can be obtained using the Franceschini-Rodighiero-Vaccari (FRV) model \cite{Franceschini08,wabside}.

We have used the recent results on neutrino mixing angles to compute flavour ratio of various neutrino flavour after oscillations. The oscillation probability for a neutrino $|\nu_\alpha\rangle$ of flavour $\alpha$ to a neutrino $|\nu_\beta\rangle$  of flavour $\beta$ 
after traversing a baseline distance $d_L$ is given by \cite{prob,prob1} 
\bea 
P_{\nu_\alpha \rightarrow \nu_\beta} &=& \delta_{\alpha\beta}
- 4\displaystyle\sum_{j>i} U_{\alpha i} U_{\beta i} U_{\alpha j} U_{\beta j}
\sin^2\left (\frac {\pi d_L} {\lambda_{ij}} \right )\,\, ,      
\label{oscprob}
\eea
where $U_{\alpha i}$ etc. in the above are the elements of the neutrino mass-flavour mixing matrix (Pontecorvo-Maki-Nakagawa-Sakata (PMNS) matrix) \cite{pmns} with $i,j$ etc. being the neutrino mass index. A neutrino $|\nu_\alpha \rangle$ of flavour $\alpha$ is related to its mass eigenstates $|\nu_i\rangle$ ($i = 1,2,3$, for three flavour case) by 
\bea
|\nu_\alpha \rangle &=& \displaystyle\sum_{i} U_{\alpha i} |\nu_i \rangle\,\, ,
\label{completeset}
\eea
If the neutrinos originate at distant blazar with the flavour ratio
$$
\varphi_{\nu_e}:\varphi_{\nu_\mu}:\varphi_{\nu_\tau} = 1:2:0 \,\, ,
$$
the flux ${\Phi^3_{\nu_\alpha}}$ on reaching the earth can be expressed as
\begin{widetext}
{\small
\begin{eqnarray}
\Phi^3_{\nu_e} &=& [ {\mid {U}_{e1} \mid}^2 (1 + {\mid {U}_{\mu1} \mid}^2 - 
{\mid {U}_{\tau1} \mid}^2 )
+ {\mid {U}_{e2} \mid}^2 (1 + {\mid {U}_{\mu2} \mid}^2  - 
{\mid {U}_{\tau2} \mid}^2 ) \nonumber\\
& &  + {\mid {U}_{e3} \mid}^2 (1 + {\mid {U}_{\mu3} \mid}^2 - 
{\mid {U}_{\tau3} \mid}^2 )]\varphi_{\nu_e}\,\,\, ,\nonumber\\
\Phi^3_{\nu_\mu} &=& [ {\mid {U}_{\mu1} \mid}^2 (1 + {\mid {U}_{\mu1} \mid}^2 - 
{\mid {U}_{\tau1} \mid}^2 ) + {\mid {U}_{\mu2} \mid}^2 
(1 + {\mid {U}_{\mu2} \mid}^2  - 
{\mid {U}_{\tau2} \mid}^2 )  \nonumber\\
& &+ {\mid {U}_{\mu3} \mid}^2 (1 + {\mid {U}_{\mu3} \mid}^2 - 
{\mid {U}_{\tau3} \mid}^2 )]\varphi_{\nu_e}\,\,\, ,\nonumber\\
\Phi^3_{\nu_\tau} &=& [ {\mid {U}_{\tau1} \mid}^2 (1 + {\mid {U}_{\mu1} \mid}^2 
- {\mid {U}_{\tau1} \mid}^2 ) + {\mid {U}_{\tau2} \mid}^2 
(1 + {\mid {U}_{\mu2} \mid}^2  - {\mid {U}_{\tau2} \mid}^2 )\nonumber\\
& & + {\mid {U}_{\tau3} \mid}^2 (1 + {\mid {U}_{\mu3} \mid}^2 - 
{\mid {U}_{\tau3} \mid}^2 )]\varphi_{\nu_e}\,\,\, . 
\label{3fmuprob}
\end{eqnarray}
}

The elements $U_{\alpha i}$ in Eq. (\ref{3fmuprob}) can be computed using
\bea
{U} &=& \left (\begin{array}{ccc}
c_{12}c_{13} & s_{12}s_{13} & s_{13} \\
-s_{12}c_{23}-c_{12}s_{23}s_{13} & c_{12}c_{23}-s_{12}s_{23}s_{13} & 
s_{23}c_{13} \\
s_{12}s_{23}-c_{12}c_{23}s_{13} & -c_{12}s_{23}-s_{12}c_{23}s_{13} & 
c_{23}c_{13}  \end{array} \right )\,\,  .
\label{pmns3f}
\eea
\end{widetext}
where $c_{{ij}(i,j = 1,2,3)} = \cos \theta_{ij}$ and $s_{{ij}(i,j = 1,2,3)} = \sin \theta_{ij}$, where $\theta_{ij}$ is the mixing angle between $i^{\rm th}$ and $j^{\rm th}$ neutrinos.

In this work we adopted the values of mixing angles 
to be $\theta_{12} = 32.96^o, \theta_{23} = 40.7^o$ and 
$\theta_{13} = 8.43^o$ \citep{pdg} for the computation. 
Hence, the flavour ratio of neutrino flux reaching at earth from the distant blazar after neutrino oscillation is evaluated to be $\Phi^3_{\nu_e}: \Phi^3_{\nu_\mu}:\Phi^3_{\nu_\tau} = 1.052:0.992:0.955$.

The corresponding muon neutrino flux reaching at the earth can be expressed as 
\begin{eqnarray}
E_{\nu}^2\frac{d\Phi_{\nu_{\mu}}}{dE_{\nu}} = \xi.\frac{V'\delta^2\Gamma_j^2}{4\pi d_L^{2}} \frac{E_{\nu}'^2}{m_e c^2}Q_{\nu,pp}'(\epsilon_{\nu}') 
\end{eqnarray}
where $\xi = 0.992/3$ is a fraction which is considered due to neutrino oscillation and $E_{\nu} = \delta E_{\nu}'/(1+z) $ \cite{Atoyan03} relates neutrino energies in the observer and co-moving jet frame respectively. If the differential flux of muon neutrinos are known then the number of expected muon neutrino event in IceCube detector in time $\tau$ can be obtained from the relation 
\begin{eqnarray}
N_{\nu_{\mu}} = \tau \int_{\epsilon_{\nu,min}}^{\epsilon_{\nu,max}} A_{eff}(\epsilon_{\nu}). \frac{d\Phi_{\nu_{\mu}}}{d\epsilon_{\nu}} d\epsilon_{\nu}
\label{event}
\end{eqnarray}
where $A_{eff}$ be the IceCube detector effective area at the declination of the blazars in the sky \cite{IceCube18b,Padovani18,Albert19}.

\section{Gamma-ray and neutrino fluxes from the Icecube detected blazars}
The gamma-ray variability time scale of all three blazars can generally be assumed as $t_{ver} \le 10^5$ s as found for TXS 0506+056 by analyzing the X-ray and gamma-ray light curves \cite{Keivani18}. The best fit spectral slope of the observed astrophysical neutrinos between 194 TeV and 7.8 PeV by IceCube observatory \cite{Halzen17,Aartsen16} suggests that the spectral index of the energy spectrum of AGN accelerated cosmic rays can be taken as $\alpha_p \sim - 2.1$ for all blazars. As the declination of the all blazars are nearly same, we have used the same $A_{eff}$ as provided by IceCube collaboration at the declination of the TXS 0506+056 in the sky for all three blazars \cite{IceCube18b,Albert19}.

\subsection{GB6 J1040+0617}
An energy of $97.4\pm 9.6$ TeV was deposited in the Icecube detector by the neutrino event, IceCube-141209A \cite{Garrappa19}. Although two neighboring FSRQs 4C+06.41 and SDSS J104039.54+061521.5 of the object GB6 J1040+0617 was found to be positionally located within the 90\% uncertainty of the well-reconstructed neutrino IceCube-141209A, they are less favored as the likely neutrino counterpart \cite{Garrappa19} because no significant high-energy gamma-ray emission was observed at the arrival of IceCube-141209A neutrino event. On the other hand, being a BL Lac object, GB6 J1040+0617 displays a bright optical flare detected by ASAS-SN observatory associated with modest gamma-ray activity at the neutrino arrival time. At the detection time of IceCube-141209A the blazar showed an increased gamma-ray activity which started a few days before the neutrino event detection and lasted for 93 days i.e, from MJD 56997 to 57090 with respect to the 9.6 years averaged flux \cite{Garrappa19}. Moreover, the blazar is located near the equatorial plane at a similar declination as TXS 0506+056 which is the sky region for which IceCube is most sensitive to high-energy neutrinos. If IceCube-141209A is astrophysical in origin, the low-synchrotron peaked gamma-ray blazar GB6 J1040+0617 appears to be a plausible neutrino source candidate based on its energetics and multiwavelength characteristics \cite{Garrappa19}. The redshift of the blazar has been recently estimated to be $z = 0.73$ \cite{Maselli15,Ahn12} and the luminosity distance of the blazar is evaluated to be $d_L \sim 4612.1$ Mpc with a consensus cosmology.

\begin{figure}[h]
  \begin{center}
  \includegraphics[width = 0.5\textwidth,height = 0.45\textwidth,angle=0]{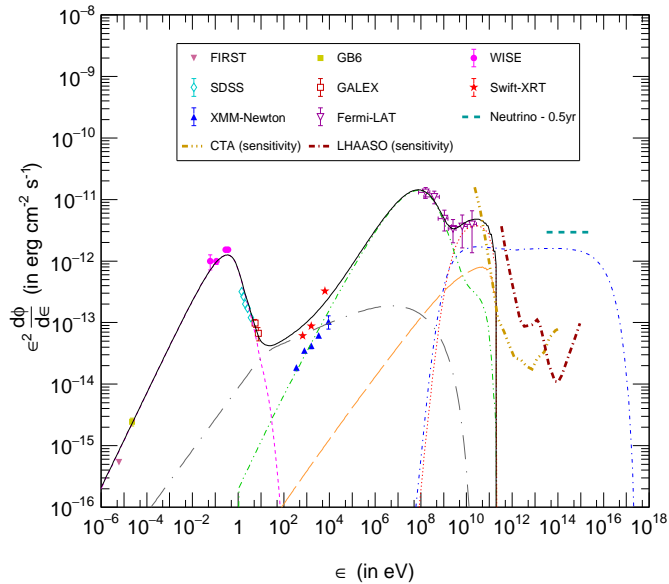}
\end{center}

  \caption{The estimated differential energy spectrum of gamma-rays and neutrinos reaching the Earth from the blazar GB6 J1040+0617. The pink small-dashed line and green long-dash-double-dotted indicate the EM spectrum due to the synchrotron emission and the inverse Compton emission  of relativistic electrons respectively. The red dotted line shows the gamma-ray flux produced from pp-interaction after $\gamma\gamma$-absorption. The gray big dash-dotted line and orange big dashed line denote the flux for synchrotron \& inverse Compton emission by electrons/positrons in EM cascade after $\gamma\gamma$-absorption. The black continuous line represents the estimated overall differential EM SED. The blue small dash-single dotted line indicates the differential muon neutrino flux at the Earth. The yellow dash-triple-dotted line and brown long dash-single dotted line denote the detection sensitivity of the CTA detector for 1000 hours and the LHAASO detector for one year respectively. The cyan long dashed line indicates the expected level and energy range of the neutrino flux reaching the Earth to produce one muon neutrino in IceCube in 0.5 years, as observed.}

\label{Fig:1}
\end{figure}

To explain the EM SED of GB6 J1040+0617 over the optical to gamma-ray energy range, we have considered the size of emission region of $R_b' = 5.2\times 10^{16}$ cm with bulk Lorentz factor of AGN jet $\Gamma_j = 21$ and Doppler boosting factor $\delta = 30$ which are strongly consistent with the size suggested from the variability, namely $R_b' \lesssim \delta c t_{ver}/(1+z) \simeq 5.2\times10^{16}$ cm assuming $t_{ver}\simeq 10^5$ s (similar to that of the blazar TXS 0506+05).

The lower energy bump of the experimental EM SED data can be explained well by the synchrotron emission of primary accelerated electron's distribution obeying a broken power law as given by Eq.~(\ref{Eq:1}) with spectral indices $\alpha_1 = 1.45$ and $\alpha_2 = 6.2$ respectively before and after the spectral break with the Lorentz factor $\gamma_b' = 1.35\times 10^{4}$. The required magnetic field and kinematic power of relativistic electrons in blazar jet to fit the observed data are found to be $B' = 0.01$ G and $L_e' = 1.5\times 10^{43}$ erg/s respectively. Also, IC scattering of primary relativistic electrons with this synchrotron photons co-moving with the AGN jet is found to produce the lower part of the high energy bump of the EM spectrum.

The higher energy bump of observed EM SED data can be reproduced well by the model as estimated using the Eq.~(\ref{eqgamflux}) and the required accelerated primary proton injection luminosity is estimated to be $L_p' = 2.7\times 10^{45}$ erg/s along with the best fitted spectral slope of $\alpha_p = - 2.1$. The cold proton number density in jet turns out to be $4.2\times 10^5$ particles/cm$^3$ under charge neutrality condition which provides sufficient targets for hadronuclear interactions with accelerated protons under the assumption of a low acceleration efficiency of electrons in AGN jet of $\chi_e \approx 10^{-3}$. The contributions of the synchrotron and IC emission of the stationary electron/positron pairs produced in EM cascade induced by protons are estimated following \cite{Bottcher13} as discussed in the previous section.  The radiative cooling time due to inverse Compton losses inversely varies with the energy density of photons in co-moving jet frame in equilibrium. For evaluating IC spectrum from cascade electrons we considered seed photons from the whole observed electromagnetic energy spectrum rather than the just synchrotron seed photons i.e. we have considered higher energy density of seed photons. This leads to smaller IC cooling time and yields higher IC and lower synchrotron fluxes from cascade electrons compared to prevailing studies. We incorporate IC process with full Kllein-Nishima formulation while implementing self-consistent formalism of Bottcher et al. (2013) \cite{Bottcher13}. 

The total jet power in the form of relativistic electron, proton kinetic energy and magnetic field evaluated as $L^k_{jet} = \Gamma_j^2 \beta_j c \pi R_b'^2\left[u_e' + u_p' + u_B' \right]$ \cite{Ansoldi18} and found out to be $1.2\times 10^{48}$ erg/s. This is consistent with the Eddington luminosity of the blazar if we assume that the system hosts a super-massive black hole of mass $M_{bh} \gtrsim 9.5\times 10^{9} M_{\odot}$, like AGN NGC 1281. During outbursts or for a collimated outflow in a jet and if the jet does not interfere with the accretion flow, the jet power may exceed the Eddington luminosity moderately, within a factor of ten \cite{Gao19,Sadowski15}.

The estimated differential gamma-ray and neutrino spectrum reaching at Earth from this blazar along with the different space and ground based observations is displayed in Fig.~\ref{Fig:1}. The X-ray data points shown in the figure are data are not contemporaneous and the displayed XMM-Newton and the Swift-XRT data were collected in 2003 May and during 2007$-$2011 respectively \cite{Garrappa19}. Note that XMM-Newton and Swift-XRT data are not mutually consistent which may be due to dynamical behavior of the source. Using the Eq.~(\ref{event}) the expected muon neutrino event in IceCube detector from the blazar is evaluated to be about $N_{\nu_{\mu}} = 0.52$ events in 32 TeV and 7.5 PeV energy range in $0.5$ years for the flaring VHE emission state with $E_{p,max}' = 20$ PeV.

\subsection{TXS 0506+056}
The IceCube detected high-energy neutrino-induced muon track IceCube-170922A was found to be positionally coincident with the known flaring $\gamma-$ray blazar, TXS 0506+056 with chance coincidence being rejected at a $3\sigma$ confidence level\cite{IceCube18a}. No additional excess of neutrinos was found from the direction of TXS 0506+056 near the time of the alert. Considering a spectral index of $-2.13$ ($-2.0$) for the spectrum of diffuse astrophysical muon neutrinos, the most probable energy of the neutrino event was estimated to be 290 TeV (311 TeV) with the 90\% C.L. lower and upper limits being 183 TeV (200 TeV) and 4.3 PeV (7.5 PeV), respectively \cite{IceCube18a, Ansoldi18}. The Fermi-LAT observations suggest that the integrated gamma-ray flux above 0.1 GeV from TXS 0506+056 elevates to the level $(5.3 \pm 0.6) \times 10^{-7}$ $cm^{-2}s^{-1}$ in the week 4 to 11 July 2017 from its averaged integrated flux of $(7.6 \pm 0.2) \times 10^{-8}$ $cm^{-2}s^{-1}$ above 0.1 GeV during 2008 to 2017. The Astro-Rivelatore Gamma a Immagini Leggero (AGILE) gamma-ray telescope observed a flux of $(5.3 \pm 2.1)\times 10^{-7}$ $cm^{-2}s^{-1}$ during 10 to 23 September 2017. Thus all the extensive follow-up observations revealed TXS 0506+056 was active in all EM bands during the period July 2017 to September 2017 whereas the source was found to be in quiescent stage most of the time based on an archival study of the time-dependent $\gamma$-ray data over the last ten years or so. The redshift of the blazar has been recently estimated to be $z = 0.3365$ \cite{Paiano18} and the luminosity distance of the blazar is evaluated to be $d_L \sim 1750$ Mpc \cite{Keivani18} with a consensus cosmology.

Here we have chosen the size of emission region of $R_b' = 5.5\times 10^{16}$ cm with bulk Lorentz factor of AGN jet $\Gamma_j = 30$ and Doppler boosting factor $\delta = 30$ which are strongly consistent with the size inferred from the variability, namely $R_b' \lesssim \delta c t_{ver}/(1+z) \simeq 6.7\times 10^{16} (\delta/30) (t_{ver}/10^5 s)$ cm \cite{Keivani18} to describe the EM SED of TXS 0506+056 over the optical to gamma-ray energy range.

\begin{figure}[h]
  \begin{center}
  \includegraphics[width = 0.5\textwidth,height = 0.45\textwidth,angle=0]{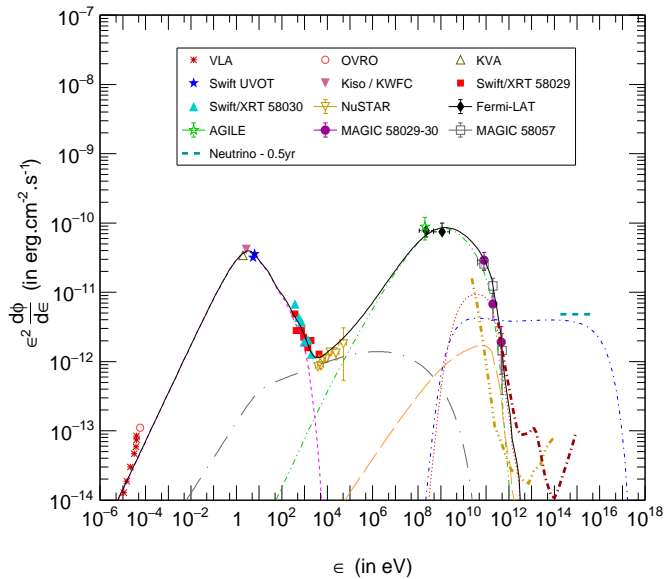}
\end{center}
  \caption{Same as Fig.~\ref{Fig:1}, but for the blazar TXS 0506+056 during the active phase. The cyan long dashed line represents the expected level and energy range of the neutrino flux at the Earth to produce one muon neutrino event in IceCube in 0.5 years, as observed.}
\label{Fig:2}
\end{figure}

We found that the synchrotron emission of primary accelerated electron's distribution obeying a broken power law can establish well the lower energy bump of the experimental EM SED data. The IC scattering of primary relativistic electrons with those synchrotron photons co-moving with the AGN jet is found to produce lower part of the high energy bump of the EM spectrum. Our proton blazar model in which gamma-rays are found to produce in interactions of relativistic protons with the ambient cold protons in the blob can explain consistently the observed higher energy bump of observed EM SED data as estimated following Eq.~(\ref{eqgamflux}). The required accelerated primary proton injection luminosity is estimated to be $L_p' = 1.65\times 10^{45}$ erg/s with the best fit spectral slope of $\alpha_p = - 2.1$ and the total kinetic jet power turns out to be $1.5\times 10^{48}$ erg/s. The synchrotron \& IC emissions of the stationary state electron/positron pairs produced in EM cascade induced by protons are evaluated found to explain well x ray data as shown in Fig.~\ref{Fig:2}. The results for this blazar was also discussed in details in \cite{Banik19} where we did not considered full Kllein-Nishima scattering cross section for computing  energy loss rate of electrons in EM cascade initiated by high energy protons. Our estimated flux from electromagnetic cascade for TXS 0506+056 is comparable with that of Liu et al. (2018) in which electromagnetic radiation absorption was also considered. The softer proton spectrum considered in the present work seems balance such effect of electromagnetic radiation absorption considered by Liu et al. (2018). Our estimated cascade synchrotron emission is almost the same to that given by Sahakyan (2018). The effect of softer primary proton spectrum considered by Sahakyan (2018) seems compensated by non-incorporation of IC process in his work. The model fitted parameters are shown in Table~\ref{table1}. 

The calculated differential gamma-ray and neutrino spectrum reaching at Earth from this blazar along with the different space and ground based observations are given in Fig.~\ref{Fig:2}. Using the Eq.~(\ref{event}) the expected muon neutrino event in IceCube detector from the blazar is found to be about $N_{\nu_{\mu}} = 0.74$ events in 200 TeV and 7.5 PeV energy range in 0.5 years for the flaring VHE emission state with $E_{p,max}' = 20$ PeV. 

\subsection{TXS 0506+056/PKS 0502+049}
Reanalyzing the historical 9.5 yr of data at the position of TXS 0506+056 Icecube collaboration reported significant evidence for a flare of $13\pm 5$ muon-neutrino events between September 2014 and March 2015 \cite{IceCube18a}. Moreover, it is also reported that observed neutrino flare has $3.5\sigma$ excess over atmospheric neutrinos in an energy range (68 percent) between 32 TeV and 3.6 PeV. But, the blazar TXS 0506+056 was appeared to be in the quiescent state in both the radio and GeV emission band at the arrival time window of such a neutrino flare \cite{IceCube18a}. The energy spectrum of detected neutrinos exhibits power law behavior with spectral slop around $-2.1$ to $-2.2$ depending on the choice of fitting window \cite{IceCube18b}. The spectral index of the accelerated proton spectrum has to be chosen accordingly.
\begin{figure}[h]
  \begin{center}
  \includegraphics[width = 0.5\textwidth,height = 0.45\textwidth,angle=0]{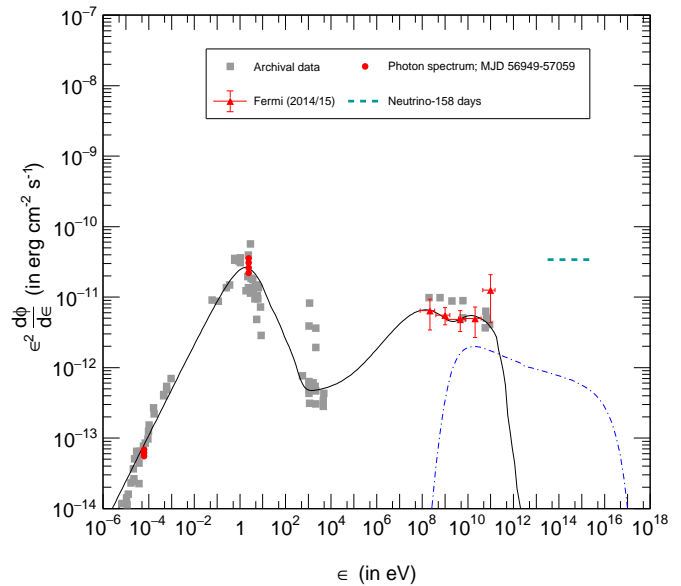}
\end{center}
  \caption{Evaluated differential energy spectrum of gamma-rays and neutrinos reaching at the Earth from the blazar TXS 0506+056 in quiescent
state (during MJD 56949$-$57059). The black continuous line and blue dash-dotted line indicates the estimated overall differential EM SED and neutrino spectrum respectively following our model. The cyan long dashed line represents the expected level and energy range of the neutrino flux reaching Earth to produce 13 muon neutrino events in IceCube in 158 rays, as observed.}
\label{Fig:3}
\end{figure}

The calculated differential gamma-ray and neutrino spectrum reaching at the Earth from TXS 0506+056 during MJD 56949-57059 along with the different space and ground based observations are given in Fig.~\ref{Fig:3}. It is found that overall differential multi wave-length EM SED of the blazar in quiescent state can be fitted using our model with the parameters like $\Gamma$, $\delta$, $\theta$ and $R_b'$ are kept fixed to that of active state of the blazar. The magnetic field has to increase slightly to $B' = 0.068$ G and other parameters are adjusted as $\alpha_1 = 1.7$, $\alpha_2 = 4.5$, $\gamma_b' = 10^{4}$, $L_e' = 9.1\times 10^{41}$ erg/s, $L_p' = 6.8\times 10^{44}$ erg/s and $L_{jet}^k = 6.1\times 10^{47}$ erg/s. The archival observed x-ray data restrict the spectral slope to $\alpha_p = - 2.2$ and the expected muon neutrino event in IceCube detector from the blazar is found to be about $N_{\nu_{\mu}} = 0.19$ events in 32 TeV and 3.6 PeV energy range in 158 days.

Thus the blazar TXS 0506+056 which is in a quiescent state during MJD 56949-57059, is unlikely to contribute to the detected neutrino flare during the same period. It is found that a bright source PKS 0502+049 blazar which is located $\sim 1.2^0$ away from TXS 0506+056, had strong GeV flares around the neutrino flare phase in 2014$-$2015. The light curve of the blazar displays two major active phases $-$ one in the periods of MJD 56860-56960 and another in the periods of MJD 57010-57120 which are partly overlapped with the 158-days box-shaped time window of IceCube neutrino flare from MJD 56937 to MJD 57096 \cite{Sahakyan19}. The redshift of the blazar has been recently evaluated to be $z = 0.954$ and the luminosity distance of the blazar is estimated to be $d_L \sim 6.4$ Gpc with a consensus cosmology \cite{Drinkwater97}.

To describe the EM SED of PKS 0502+049 over the optical to gamma-ray energy range during MJD 56860-56960, we have considered the size of emission region of $R_b' = 6\times10^{16}$ cm with bulk Lorentz factor of AGN jet $\Gamma_j = 30$ and Doppler boosting factor $\delta = 40$. The size of emission region $R_b'$ is strongly consistent with the size inferred from the variability, namely $R_b' \lesssim \delta c t_{ver}/(1+z) \simeq 6.14\times10^{16}$ cm assuming $t_{ver}\simeq 10^5$ s (similar to the blazar TXS 0506+05).

\begin{figure}[h]
  \begin{center}
  \includegraphics[width = 0.5\textwidth,height = 0.45\textwidth,angle=0]{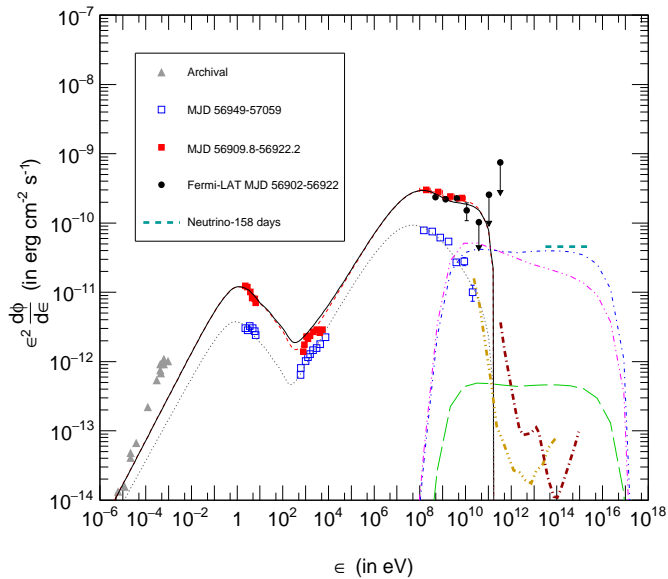}
\end{center}
  \caption{The estimated differential energy spectrum of gamma-rays and neutrinos reaching the Earth from the blazar PKS 0502+049. The black continuous line (or, red small-dashed line) and blue small-dash-dotted line (or, pink small dash-double-dotted line) indicate the estimated overall differential EM SED and the differential muon neutrino flux at the Earth following our model using $\alpha_p = - 2.1$ (or, $- 2.2$) during first active phase in MJD  56909.8$-$56922.2. The gray dotted line and the green very-long dashed line represent the same with $\alpha_p = - 2.1$ but for the quiescent state of the blazar during MJD 56949$-$57059. The yellow dash-triple-dotted line and brown long-dash–single-dotted line indicate the detection sensitivity of the CTA detector for 1000 hours and the LHAASO detector for one year, respectively. The cyan long dashed line represents the expected level and energy range of the neutrino flux reaching the Earth to produce 13 muon neutrino events in IceCube in 158 days, as observed.}
\label{Fig:4}
\end{figure}

During the first active phase, the synchrotron emission of primary accelerated electron's distribution is found to produce well the lower energy bump of the experimental EM SED data of the blazar PKS 0502+049. The required magnetic field and kinematic power of relativistic electrons in blazar jet as given by Eq.~(\ref{Eq:2}) are $B' = 0.023$ G and $L_e' = 1.7\times 10^{43}$ erg/s respectively for reproduction of the em spectrum. The IC scattering of primary relativistic electrons with this synchrotron photons co-moving with the AGN jet are found to produce the lower part of the high energy bump of the EM spectrum. 

Following the model as estimated using Eq.~(\ref{eqgamflux}), the observed higher energy bump of observed EM SED data during first active phase can be reproduced well and the required accelerated primary proton injection luminosity is estimated to be $L_p' = 9.2\times 10^{45}$ erg/s with the spectral slope of $\alpha_p = - 2.1$ to $-2.2$. Here we have taken acceleration efficiency of the electrons in AGN jet as $\chi_e \approx 10^{-3}$ and the cold proton number density in jet turns out to be $1.4\times 10^6$ particles/cm$^3$ under charge neutrality condition which provides sufficient targets for hadronuclear interactions with accelerated protons. The expected muon neutrino event in IceCube detector from the blazar is found to be about $N_{\nu_{\mu}} = 10.85$ and $5.2$ events using $\alpha_p = - 2.1$ and $-2.2$ respectively in 32 TeV and 3.6 PeV energy range in 158 days for the flaring VHE emission state with $E_{p,max}' = 20$ PeV as estimated using Eq.~(\ref{event}).

The total jet power in the form of relativistic electron, proton kinetic energy and magnetic field is estimated as $L^k_{jet} = 8.3\times 10^{48}$ erg/s which is about 84 times higher then the Eddington luminosity of the blazar PKS 0502+049 ($L_{edd}\simeq 9.8\times 10^{46}$ erg/s considering the black hole mass of $M_{bh} \simeq 7.53\times 10^{8} M_{\odot}$ \cite{Oshlack02}). The synchrotron and cascade emissions of the electrons/positrons pairs produced in EM cascade initiated by primary accelerated protons can explain the observed x-ray data.

On the other hand, the quiescent state of the blazar during MJD 56949-57059 found to be of leptonic origin where the observed low- and high-energy bump of observed EM SED data can be explained well by the synchrotron emission and IC scattering of primary accelerated electron's distribution of injection luminosity $L_e' = 1.3\times 10^{43}$ erg/s with $B' = 0.017$ G. Because of harder spectral slope at GeV energies of observed EM SED, the second active phase during MJD 57010-57120 may also be originated in leptonic mechanism as suggested by Sahakyan (2019) \cite{Sahakyan19}. The gamma-ray and neutrino production via $pp$ interaction may be sub-dominant and hence emission of neutrinos is not significant during this phase. The expected muon neutrino event in IceCube detector from the blazar is found to be about $N_{\nu_{\mu}} = 0.13$ with $L_p' = 8\times 10^{43}$ erg/s and $\alpha_p = - 2.1$ in 32 TeV and 3.6 PeV energy range in 158 days with $E_{p,max}' = 20$ PeV. The calculated differential gamma-ray and neutrino spectrum reaching at Earth from this blazar along with the different space and ground based observations is shown in Fig.~\ref{Fig:4}. 

\begin{table*}[t]
  \begin{center}
    \caption{Model fitting parameters for TXS 0506+056, PKS 0502+049 and GB6 J1040+0617  according to proton blazar model.}
    \label{table1}
{\setlength{\tabcolsep}{1.4em}
\renewcommand{\arraystretch}{1.4} 
    \begin{tabular}{m{2.3cm}|cc|c|c}
      \toprule
      Parameters & \multicolumn{2}{c|}{TXS 0506+056} &  PKS 0502+049 & GB6 J1040+0617 \\ \cline{2-5}
                 &    Active & Quiescent             & Active &  Active\\         \clineB{1-5}{1.5}
      $\delta$   &  $30$       &  $30$       &        40            &       30 \\
      $\Gamma_j$ &   $30$      &  $30$       &        30            &      21 \\
      $\theta$   &   $1.9^{0}$ &  $1.9^{0}$  &   $1.35^{0}$         &  $1.7^{0}$ \\
       $z$       &   $0.3365$  & $0.3365$    &  $0.954$             &  $0.73$ \\
 $R_b'$  (in cm) &  $5.5\times 10^{16}$ & $5.5\times 10^{16}$ &   $6\times 10^{16}$   &  $5.2\times 10^{16}$\\         
    $B$ (in G)   &  $0.034$    &$0.068$      &  $0.023$       &  $0.01$     \\
   $\alpha_1$    &   $- 1.6$   & $-1.7$      &   $- 1.7$      &   $- 1.45$ \\ 
   $\alpha_2$    &   $- 4.1$   & $-4.5$      &     $- 3.8$    &    $- 6.2$\\
    $\gamma_b'$  &  $1.8\times 10^{4}$   & $10^{4}$             & $1.3\times 10^{4}$  &   $1.35\times 10^{4}$ \\
 $\gamma_{e,min}'$    &        $1$         &  $1$  &         $1$          &  $1$\\
 $\gamma_{e,max}'$    &  $3\times 10^{5}$  &  $10^5$   &$1.4\times 10^{5}$ &  $10^{5}$     \\
 $L_e'$ (in erg/s)          & $2.3\times 10^{42}$   & $9.1\times 10^{41}$  &$1.7\times 10^{43}$ &   $1.5\times 10^{43}$  \\ 
 $n_H$  (in cm$^{-3}$)      &  $1.1\times 10^5$     & $10^5$               & $1.4\times 10^6$ &     $4.2\times 10^5$     \\
 $\alpha_p$                 &   $- 2.1$             &  $- 2.2$             &  $- 2.1$     &    $- 2.1$\\
 $E_{p,max}'$ (in eV)       &  $ 2\times 10^{16}$   & $ 2\times 10^{16}$   &  $ 2\times 10^{16}$    &        $2\times 10^{16}$      \\ 
       $L_p'$ (in erg/s)    &  $1.65\times 10^{45}$ & $6.8\times 10^{44}$  & $9.2\times 10^{45}$ &       $2.7\times 10^{45}$ \\
   $L^k_{jet}$ (in erg/s)   &  $ 1.5\times 10^{48}$ & $6.1\times 10^{47}$  &$ 8.3\times 10^{48}$     &     $ 1.2\times 10^{48}$\\
          $N_{\nu_{\mu}}$   &  $0.74$               & $0.19$               & $10.85$ &       $0.52$\\ \hline   \hline

    \end{tabular}}\quad
  \end{center}
\end{table*}

\section{Discussion}
Observation of high energy neutrinos along with gamma-rays from flaring blazars suggest a hadronic mechanism as their origin in the blazar jets. Although AGN jet composition is not conclusively known so far, the general convention is that there is a dense radiation target within the jet unless it is of an external origin \cite{Sahakyan19}. The photomeson reaction ($p\gamma$) is more widely used to explain the emission from blazars. To describe the spectral energy distribution (SED) and the neutrino event, the conventional one-zone models offer too low neutrino rates in combination with excessively high neutrino energies or maintained super-Eddington injection luminosities. The current theoretical study generally suggests that the geometry of the radiation zone must be more complex, involving an external radiation field boosted into the jet frame, either thermal \cite{Keivani18} or non-thermal with structured jet modeling \cite{Ansoldi18} or a compact radiation core with high photohadronic interaction rates \cite{Gao19}. By adopting such complex geometry of radiation zone, one may interpret the neutrino events from TXS 0506+056 (during the year 2017) and GB6 J1040+0617 but the observation of flaring events during the period 2014-15 from the direction of TXS 0506+056/PKS 0502+049 cannot be explained through lepto-hadronic models \cite{Rodrigues19, Sahakyan19}.    

In inelastic hadronic $pp$ interaction scenario requires high thermal plasma density but offers high neutrino rates in combination with neutrino energy range as detected by IceCube observatory. The scenario is effectively understood in cloud-jet interaction scenario where the presence of clouds in the vicinity of the super-massive black hole (SMBH) that provides targets for inelastic $pp$ collisions once they enter the jet. However, the presence of broadline region (BLR) clouds in the vicinity of the SMBH for TXS 0506+056 is doubtful due to the non-detection of the BLR emission from TXS 0506+056, GB6 J1040+0617 and other BL Lac objects. Consequently the hadronuclear (pp) interaction interpretation like relativistic jet meets with high density cloud is unlikely a common scenario for neutrino production in all IceCube blazars. 

In the original proton blazar model \cite{Mucke01} high energy gamma-rays are interpreted through synchrotron radiation by high energy protons in a strong magnetic field environment. However, the gamma-ray spectrum of the IceCube blazars can not be modeled with the proton synchrotron radiation due to low magnetic field strength of the source as obtained from the fitting of low energy hump of SED. In the present cases, the photomeson ($p\gamma$) interaction is also found inefficient due to the low amplitude of the target synchrotron photon field. 
Instead, the inspired proton blazar model as elaborately discussed in Banik \& Bhadra (2019) \cite{Banik19} appears to be more appropriate as the high energy gamma-rays are produced along with relatively higher neutrino event rate in interactions of relativistic protons with the ambient cold protons in the blob of AGN jet. In the framework of the proton blazar model, our findings suggest that relative contributions to the total jet power of cold protons, accelerated protons, magnetic field, and accelerated electrons obtained based on charge neutrality can describe consistently both the low- and high-energy bump of the multiwavelength EM SED and also the detected neutrino events IceCube-170922A and IceCube-141209A from the flaring blazar TXS 0506+056 and GB6 J1040+0617 respectively. On the other hand, the emission of 13 muon-neutrino events observed in 2014-2015 from the direction of the blazar TXS 0506+056 can not be explained by any existing models assuming the correlation the neutrino flare with TXS 0506+056 which was found to be in the quiescent state at the arrival time window of such a neutrino flare. The number of neutrino events which was estimated is conservative because of possible contribution from interactions of tau-neutrinos that induce muons with a branching ratio of 17.7\%  \cite{Ansoldi18}. With this consideration, the total muon like neutrino events can be estimated as $N_{\mu}^{like} = N_{\mu} + 17.7\% \times \frac{0.955}{0.992} N_{\mu}$ and found out to be $0.61$, $0.86$ and $12.7$ from GB6 J1040+0617, TXS 0506+056 in 2017 flare and PKS 0502+049 in 2014-2015 flare respectively for cosmic ray spectral index $\alpha_p = -2.1$.
We found that the nearby flaring blazar PKS 0502+049 can effectively contribute 13 neutrino events depending on the cosmic ray spectral slope to the IceCube reported neutrino flare in 158-days during its first active phase (MJD 56860-56960). The second active phase of PKS 0502+049 during MJD 57010-57120 is supposed to be leptonic in origin as suggested by Sahakyan (2019) \cite{Sahakyan19} and consequently, no neutrino emission is expected during this stage. We find that the maximum energy ($E_p = \delta E_p'/(1+z)$) of the cosmic ray particle achievable in the blazars, namely TXS 0506+056, PKS 0502+049 and GB6 J1040+0617 are $4.5\times 10^{17}$ eV, $4.1\times 10^{17}$ eV and $3.5\times 10^{17}$ eV respectively in the observer frame and are required to explain consistently the observed gamma-ray and neutrino signal from the sources. 
The model fitting parameters to match the EM SED, as well as muon neutrino events from each of the blazars considered here, is shown in Table~\ref{table1}.

\section{Conclusion}
The coincident detection of the neutrino event, IceCube-170922A, 13 muon-neutrino events observed in 2014-2015 and IceCube-141209A by IceCube observatory with the gamma-ray flaring blazar, TXS 0506+056, PKS 0502+049 and GB6 J1040+0617 respectively provide support to the acceleration of cosmic rays in the blazar jet in diffusive shock acceleration process \cite{IceCube18a}. The photomeson reaction ($p\gamma$) is more widely used to explain the emission from blazars with major shortcomings of either low neutrino rates or complex geometry of the blazar jet required. More importantly, the lepto-hadronic model can not reproduce the neutrino flaring events from the direction of TXS 0506+056/PKS 0502+049 during the period 2014-15. On the other hand, the cloud-jet interaction scenario seems unlikely to be a common scenario for neutrino production in all IceCube blazars due to the absence of the broadline emission in the optical spectra of the sources. 

In the framework of the proton blazar model, our findings suggest that both the low- and high-energy bump of the multiwavelength EM SED and also the observed neutrino events from the corresponding blazars can be explained consistently with the relative contributions to the total jet power of cold protons, accelerated protons, accelerated electrons and magnetic field obtained based on charge neutrality. The (present) model predicted flux is in minor tension with the observed x-ray data which has been found in earlier p$\gamma$ and hadronuclear models as well. Such a minor discrepancy may be due to the absorption of x-rays (by mainly hydrogen column) while reaching the solar system from the source. The nearby flaring blazar PKS 0502+049 of the TXS 0506+056 is found to contribute mostly to the neutrino flare observed during 2014-2015 which can not be described by any model from TXS 0506+056 as origin \cite{Rodrigues19}.

Our findings suggest that the maximum energy of the cosmic ray particle achievable in the blazars is nearly one order less than the ankle energy of the cosmic ray energy spectrum in the observer frame and is required to explain consistently the observed gamma-ray and neutrino signal from the IceCube sources. The upcoming gamma-ray experiments like CTA \cite{Ong17} and LHAASO \cite{Liu17}, which are very sensitive up to 100 TeV energies, may provide clearer evidence of the physical origin of gamma-rays and maximum achievable energy of cosmic rays in AGN jets if more such events are detected from other blazars in upcoming years.

\section*{Acknowledgments}
The authors would like to thank an anonymous reviewer for insightful comments and very useful suggestions that helped us to improve and correct the manuscript. One of the authors (M.P.) thanks the DST-INSPIRE fellowship grant (DST/INSPIRE/ FELLOWSHIP/IF160004) by DST, Govt. of India. AB acknowledges the financial support from SERB (DST), Govt. of India vide approval number CRG/2019/004944.


\end{document}